\documentclass[lettersize,journal]{IEEEtran}

\usepackage[table,xcdraw]{xcolor}

\usepackage{scalerel}
\usepackage{tikz}
\usetikzlibrary{svg.path}

\definecolor{orcidlogocol}{HTML}{A6CE39}
\tikzset{
  orcidlogo/.pic={
    \fill[orcidlogocol] svg{M256,128c0,70.7-57.3,128-128,128C57.3,256,0,198.7,0,128C0,57.3,57.3,0,128,0C198.7,0,256,57.3,256,128z};
    \fill[white] svg{M86.3,186.2H70.9V79.1h15.4v48.4V186.2z}
                 svg{M108.9,79.1h41.6c39.6,0,57,28.3,57,53.6c0,27.5-21.5,53.6-56.8,53.6h-41.8V79.1z M124.3,172.4h24.5c34.9,0,42.9-26.5,42.9-39.7c0-21.5-13.7-39.7-43.7-39.7h-23.7V172.4z}
                 svg{M88.7,56.8c0,5.5-4.5,10.1-10.1,10.1c-5.6,0-10.1-4.6-10.1-10.1c0-5.6,4.5-10.1,10.1-10.1C84.2,46.7,88.7,51.3,88.7,56.8z};
  }
}

\newcommand\orcidicon[1]{\href{https://orcid.org/#1}{\mbox{\scalerel*{
\begin{tikzpicture}[yscale=-1,transform shape]
\pic{orcidlogo};
\end{tikzpicture}
}{|}}}}

\usepackage[numbers]{natbib}
\usepackage{graphicx}
\usepackage{listings}
\usepackage{url}
\usepackage{xurl}
\usepackage{subcaption}
\usepackage{soul}

\usepackage{hyperref}

\usepackage{multirow}

\usepackage{booktabs}

\title{Zero-Trust Artificial Intelligence Model Security Based on Moving Target Defense and Content Disarm and Reconstruction}

\author{Daniel Gilkarov \orcidicon{0009-0008-9274-802X},
\IEEEmembership{Student Member, IEEE},\\
Ran Dubin \orcidicon{0000-0002-2055-2211},
\IEEEmembership{Member, IEEE}
\thanks{Daniel Gilkarov is with the Department of Computer Science and Ariel Cyber Innovation Center, Ariel University, Ariel, Israel. e-mail: (daniel.gilkarov1@msmail.ariel.ac.il).}
\thanks{Ran Dubin is with the Department of Computer and Software Engineering and Ariel Cyber Innovation Center, Ariel University, Ariel, Israel. e-mail: (rand@ariel.ac.il).}
\thanks{Manuscript received ????; revised ????}}

\begin{document}

\maketitle

\begin{abstract}
This paper examines the challenges in distributing AI models through model zoos and file transfer mechanisms. Despite advancements in security measures, vulnerabilities persist, necessitating a multi-layered approach to mitigate risks effectively. The physical security of model files is critical, requiring stringent access controls and attack prevention solutions. This paper proposes a novel solution architecture composed of two prevention approaches. The first is Content Disarm and Reconstruction (CDR), which focuses on disarming serialization attacks that enable attackers to run malicious code as soon as the model is loaded. The second is protecting the model architecture and weights from attacks by using Moving Target Defense (MTD), alerting the model structure, and providing verification steps to detect such attacks. The paper focuses on the highly exploitable Pickle and PyTorch file formats. It demonstrates a 100\% disarm rate while validated against known AI model repositories and actual malware attacks from the HuggingFace model zoo.
\end{abstract}

\section{Introduction}
The swift evolution of Artificial Intelligence (AI) technology has made it a top priority for cybercriminals looking to obtain confidential information and intellectual property. These malicious individuals may try to exploit AI systems for their own gain, using specialized tactics alongside conventional IT methods. Given the broad spectrum of potential attack strategies, safeguards must be extensive. Experienced attackers frequently employ a combination of techniques to execute more intricate operations, which can render layered defenses ineffective.

While adversarial AI model security \cite{chen2024zddr,nguyen2023physical}, privacy \cite{yao2024survey} and operational security aspects of AI receive much attention \cite{NSA_AI_MODEL_sec,mitre_atlas}, it's equally important to address the physical file security aspects of AI models. Various malware attacks can target AI model files, and if successful, they can have severe consequences for businesses, users, and AI-driven systems.

\begin{figure}[htp]
    \includegraphics[width=8cm]{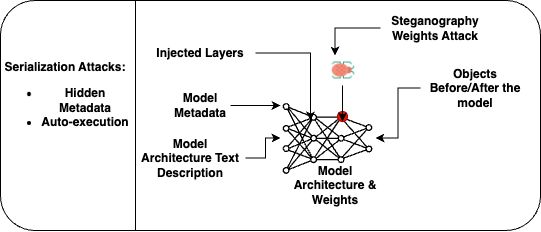}
    \caption{Simplified AI model file architecture}
    \label{fig:model_architecture}
\end{figure}

Fig. \ref{fig:model_architecture} shows the simplified AI model architecture. The AI model file involves serialization that packs the model weights and architecture. Malicious code can be hidden inside the serialization and is known to be exploitable and act as code execution \cite{hiddenlayer,models_are_code}. Inside the model, we have the architecture metadata (in some), model weights, and optional metadata such as labels and notes.

The characteristics of AI models make them an excellent target for hackers since their sizes and structure simplify hiding malicious content inside the model metadata and weights or using known vulnerable serialization to run code. Marco Slaviero \cite{Pickle_serialization_vul} demonstrated in 2011 that an attacker can abuse Pickle serialization to execute arbitrary code or system operations. Since then, the file format has been known to be insecure, but it is still widely used, for example, in PyTorch model serialization \cite{pytorchhub}. According to \cite{malhug}, about 55\% of the models in HuggingFace repositories are Pickle-based (PyTorch, Joblib, and Dill.).

Organizations may experiment with various models shared through social media and use popular models Zoo (model sharing platforms) such as  Hugging Face~\cite{Huggingfacecite}, Pytorch Hub~\cite{pytorchhub}, and others). For example, at the time of this paper writing, Hugging face model Zoo shared a fine-tuned model named "jonatasgrosman/wav2vec2-large-xlsr-53-english" that reached over almost 57 million downloads in April 2024 \cite{hftopmodel} and the recent Mistral-7B-v0.1 Large Language Model (LLM) has reached 74,356 downloads in just two days. This means that the community that uses AI models is extensive and up-to-date, and this can cause malicious model propagation to spread extremely fast. Furthermore, the models can be downloaded automatically through a simple one-line code, making generating attacks even more effortless.

Currently, the Hugging Face model Zoo uses two security mechanisms: Pickle~\cite{pickle_standard} malicious serialization detection~ \cite{Huggingfacepickle}, and anti-virus detection based on ClamAV \cite{ClamAV} to detect malicious models. However, they do not support detecting attacks targeting the model weights, may miss serialization attacks \cite{pickle-attack}, and metadata malware hiding attacks. A recent report  \cite{jfrog_ai_model} emphasizes that Hugging Face conducts scans on Pickle models; however, it does not outright block or restrict them from being downloaded but marks them as “unsafe”. As a result, an additional security layer is needed to prevent the attack during deployment. Furthermore, their research concludes that 95\% of the attacks target Pickle serialization while the rest are TensorFlow serialization.

The National Security Agency (NSA) recently released a Cybersecurity Information Sheet (CSI) \cite{NSA_AI_MODEL_sec}. The CSI is intended to support National Security System owners and Defense Industrial Base companies deploying and operating AI systems designed and developed by an external entity. The CSI illustrates the steps to protect the deployment and two main points directly related to this work. 
The first is to validate and test the AI model. However, this is a tricky problem since current detection may fail, as will be explained later. The second is protecting AI model weights. The NSA CSI suggests the following protection steps:
\begin{itemize}
    \item Harden interfaces for accessing model weights to increase the effort it would take for an adversary to exfiltrate the weights.
    \begin{itemize}
    \item Implement hardware protections for model weight storage, disable unnecessary hardware communication capabilities, and protect against emanation or side-channel techniques.
    \item Aggressively isolate weight storage. For example, store model weights in a protected storage vault or enclave
    \end{itemize}
\end{itemize}

The Open Web Application Security Project (OWASP) has been at the forefront of web application security for years. Recognizing the burgeoning significance of machine learning, OWASP introduced the "Top 10 Machine Learning Risks" \cite{OWSAPtop10ml}, a compilation that categorizes and outlines the most prevalent threats targeting AI systems, specifically about the content of the model and the threats around it. Examples of the discussed attacks are inference attacks and poisoning attacks. Of all the discussed threats, the AI Supply Chain Attacks are the most relevant to this work, while the other attacks are related to the content and integrity of the model. In the  AI supply chain attack use case, an attacker modifies or replaces a machine learning library or model used by a system. This can also include the data associated with the machine learning models. MITRE Adversarial Threat Landscape for Artificial-Intelligence Systems (ATLAS™) \cite{ATLAS} is a comprehensive knowledge base. It's designed to document adversary tactics, techniques, and case studies pertinent to machine learning (ML) systems. These details are amassed from real-world observations, demonstrations by ML red teams and security groups, and academic research. In terms of this research, ATLAS helps to understand the attack surface, but it is not a cyber solution. This work is positioned in MITRE ATLAS as a solution against user execution of unsafe ML Artifacts (AML.T0011.000)  \cite{mitre_atlas_unsafeml}.

While growing, the literature on physical AI model security is still in its early stages \cite{dubin2023disarmingAccess, pickle-attack, PengZhou, Pickle_serialization_vul}. As the adoption of AI and Machine Learning (ML) technologies continues to rise, the need for robust security measures becomes increasingly critical. This work focuses on providing a secure solution to download and transfer AI models.

To ensure full protection for AI model distribution, whether in a file format or in a supply chain where part of the software is AI models, we have proposed two solutions that can be used together or separately. The first is a novel restricted-unpickler \cite{PickleRestriction} implementation for developing Content Disarm and Reconstruction (CDR) methods. We call our solution "safe-unpickler", and the process it executes "safe-unpickling". The second is model weight separation based on the Moving Target Defense (MTD) of model weights. Our MTD also provides verification and authentication mechanisms to guarantee the model's validity. 

The paper is organized as follows: Section \ref{Contribution} summarizes the paper's contribution. Section \ref{Related Work} reviews the related work. Section \ref{Pickle_Serialization} outlines the Pickle serialization structure and highlights the various attack vectors associated with the serialization. Section \ref{Architecture} presents the safe-unpickler, CDR, and MTD solution architecture. Section \ref{Dataset} offers a detailed description of our dataset.
In Section \ref{Evaluation}, we present our research evaluation and validation methodology. Subsequently, in Section \ref{Limitations}, we discuss our limitations. Finally, Section \ref{Conclusions} provides our conclusions and suggests future directions for research.

\section{Contribution}
\label{Contribution}

This work reviews the state-of-the-art physical Artificial Intelligence (AI) model security solutions. For the first time as far as we know, we suggest two novel zero-trust solutions:
\begin{itemize}
\item \textbf{Moving Target Defense (MTD) for AI Model Security:} A MTD solution designed to bolster the robustness and security of AI models against potential threats. This approach ensures protection throughout the entire model life-cycle, from its initial training phase to its final distribution. Consequently, it provides a strong defense against physical file-based attacks on the model weights and prevents unauthorized use of the trained model by obfuscating the model weights.

\item \textbf{AI Model Serialization Attacks Prevention and CDR:} An innovative method to neutralize Model Serialization Attacks (MSA) without needing detection mechanisms and load potentially risky models using our novel safe-unpickler-based CDR method.
\end{itemize}

\section{Related Work}
\label{Related Work}
We will begin our review by discussing Model Serialization Attack (MSA) and Content Disarm and Reconstruction (CDR) related works \ref{CDR Related work}, followed by a presentation of MTD-related research \ref{MTD Related work}.

\subsection{MSA prevention Related works}
\label{CDR Related work}

Pickle security research primarily concentrates on devising intricate attack strategies ~\cite{Pickle_attack_wild, PengZhou,huang2022pain} and developing robust protective measures \cite{PickleRestriction}. These attacks will be benchmarks for evaluating our proposed solution compared to existing protection mechanisms. 

\cite{Pickle_attack_wild} reviews the latest three malware attacks exploiting Pickle in the wild. Hung et al. \cite{huang2022pain} investigate the factors that cause failure in Pickle Unpickler and proposed Pikora \cite{Pikora}, a small compiler to compile Pickle scripts to opcodes to evaluate the robustness of restricted-unpicklers. Hung et al. \cite{huang2022pain} evaluated nine restricted-unpickler implementations from GitHub \cite{petastorm} and concluded that if the \textit{module} and \textit{name} parameters of \textit{find\_class} are restricted at the same time, the file load is likely to be safe. Pickle-Fuzz \cite{Pickle-Fuzz} uses fuzzing to detect new Pickle Denial Of Service (DOS) attacks while loading a Pickle file. However, those attacks slow the load, crash Python, or cause loading failures and are much less dangerous compared to the remote code execution this work is focused on. Boyan Milanov proposed Sleepy Pickle \cite{SleepyPickle}, a combined Pickle attack based on modifying model weights (handled in this work by MTD) and execution using the aforementioned Pickle deserialization exploit.
Fickling \cite{Fickling} is a decompiler, static analyzer, and bytecode rewriter for Python Pickle object serializations. ModelScan \cite{modelscan} is an AI model scanner focusing on serialization attacks and supports Pickle and other file formats. This work uses Fickling to validate Pickle models.

Peng Zhou~\cite{PengZhou} showed that Hugging Face Pickle security is even worse and suggested four new methods to bypass their alerts.
\begin{enumerate}
    \item \textbf{Long names:} Using a long file name bypasses the scanning.
    \item \textbf{Code reuse}: Use vulnerable libraries using Pickle to load the next phases of the attack indirectly. This relate also to the Config abuse.
    \item \textbf{Config abuse}: In this scenario, the user downloads a model from Hugging Face. The imports used are in the allowlist and shown in grey. After loading the model, the user parses the configuration file, which contains a malicious element that points to additional content downloaded from another Hugging Face repository identified as malicious. However, the user is unaware of this since the download is done automatically without the user's knowledge or approval.
    \item \textbf{Format encode}: using base64 decoding to hide malicious capabilities inside the PyTorch data section.
\end{enumerate}
Peng Zhou~\cite{PengZhou} work emphasizes that Pickle attacks are direct runtime attacks but can also work indirectly. The conclusion is that the Pickle and the libraries it calls and additional files distributed with the Pickle and remote calling (indirect) files should be analyzed. In this work, the safe-unpickling process considers these risks a concern, and organizations should avoid automatically downloading models and artifacts. Critical infrastructures and enterprises should validate and investigate the framework dependencies and supply chains to avoid hidden risks.

On the protection side, Pickle suggested Restricting Globals~\cite{PickleRestriction}, an allowlist approach based on inherited \textit{Pickle.Unpickler} and customized \textit{find\_class} method. All "restricted-unpickler" implementations we've come across do so by overloading "find\_class", \cite{huang2022pain} enforces this. We call the previous "restricted-unpickler" implementations "find\_class" restricted-unpicklers. Uber implemented their own restricted-unpickler \cite{petastorm} that defines a list of modules that the developer considers safe, however, Hung et al. \cite{huang2022pain} reported that a malicious user can still get the functions \textit{builtins.eval} or \textit{builtins.exec} to execute arbitrary code and bypass their restrictions. In this work, we compare our method to two restricted-unpickler implementations cited by \cite{huang2022pain}; The first is part of the FederatedAI/Fate GitHub repository \cite{fate}, and we use it since it was specifically designed to load PyTorch models. The second is part of the Ultimaker/Uranium GitHub repository \cite{uranium}, and we use it since \cite{huang2022pain} denoted it as the most robust of the restricted-unpickler implementations explored. 
The "find\_class" restricted-unpicklers all share a fault by design, they can't load a pickle that contains GLOBAL operations that import prohibited libraries, this doesn't allow CDR because these models can't be loaded at all.

Our proposed method aims to refine the "find\_class" restricted-unpickler approach by incorporating the previously discussed methods \cite{huang2022pain} and augmenting the technique by overloading the "load\_reduce" function of pickle.Unpickler. Our approach is more flexible and can safely load sub-objects that are safe within partially compromised pickled objects, as opposed to the "find\_class" restricted-unpicklers that fail to load an object if a forbidden function/library is requested. Moreover, our approach allows allowing/forbidding calls based on the parameters of the call, which allows for more fine-grained control, whereas "find\_class" restricted-unpicklers only control the safe-unpickling process from a global, import perspective.

The proposed MSA prevention technique was developed to address a growing concern in organizations that seek to utilize pre-trained AI models. The safe-unpickling engine can be used to purify models before they enter enterprise systems. This process is widely known as Content Disarm and Reconstruction (CDR).

CDR is a file security method based on zero-trust \cite{gauravfuture} principles. It does not rely on detection and is applied to every file, regardless of whether it is known to be malicious or not. This method is gaining popularity across various sectors, such as Industrial Control Systems (ICS), file upload protection, and email security
\cite{belkind2023open,dubin2023contentpdf,dubin2024contentole,dubin2023contentrtf, dubin2023disarmingAccess,DocBleach,Disinfect:pdfCDR}.
CDR's effectiveness stems from its deep understanding of file format specifications and its ability to pinpoint potential vulnerabilities. Therefore, CDR must comprehensively address each supported file type for optimal security and consistently validate its efficacy. The most relevant work for AI model CDR focuses on defending against steganography attacks \cite{dubin2023disarmingAccess} and suggests two methods. One is based on random bit modification, and the other on model weight quantization. This work provides a way to perform CDR on AI models to address the threat of MSA specifically.

Many related works focused on protecting AI models from malicious inputs such as prompt injection \cite{greshake2023not}, detecting adversarial attacks \cite{chen2024zddr} and validating their outputs from security and privacy aspects \cite{yao2024survey}, gender, bias, and stereotypes \cite{kotek2023gender}. However, this work focuses on the physical security of the AI model and the ability to prevent the exploitation of such models.

\subsection{MTD Related work}
\label{MTD Related work}
Moving Target Defense (MTD)~\cite{cho2020toward} is a proactive cybersecurity strategy that increases the complexity and cost for attackers by continuously changing the attack surface. Unlike traditional defense mechanisms, MTD introduces variability and unpredictability into the system, making it more challenging for attackers to exploit vulnerabilities. MTD encompasses techniques like IP address hopping \cite{heydari2018moving}, port rotation \cite{heydari2018moving}, virtual machine migration \cite{azab2016migrate}, code diversification \cite{styugin2016new}, software obfuscation \cite{banescu2018tutorial}, memory randomization \cite{mordehai2017method}, and data randomization \cite{evans2011effectiveness}. It is helpful in network and software security and represents a paradigm shift in cybersecurity by emphasizing dynamic and adaptive defense mechanisms. Our AI model MTD uses model weight randomization and reconstruction to prevent all model weight attacks and report an attempt of an attack.

\section{Pickle Serialization}
\label{Pickle_Serialization}
Before the distribution or deployment of a trained machine learning (ML) model, the model must undergo a process known as serialization. This process entails the conversion of the model into a byte stream format, facilitating its storage, transmission, and subsequent loading. Serialization represents a conventional method applicable to diverse data structures and objects. Among the prevalent generic serialization formats are CSV, JSON, XML, and Google Protobuf \cite{protobuf}. In this paper, we will focus on the most common one, named Pickle \cite{pickle_standard}. While Pickle was not designed explicitly for machine learning purposes, and it is known to be vulnerable from 2011 \cite{Pickle_serialization_vul}, it is still the most common serialization format and the default serialization for PyTorch \cite{pytorchhub}.

Pickle converts Python objects into a binary format that can be stored or transmitted efficiently. This binary format is platform-independent, meaning it can be read and written by any Python interpreter, regardless of the operating system or architecture. Pickle relies on a tiny virtual machine called Pickle Machine (PM). PM Reads opcodes and arguments from the stream, starting at byte 0, Processes them, alters the stack, and repeats until the end of the stream. The procedure returns the top of the stack as the deserialized object "STOP" is encountered.

\begin{figure*}
    \centering
    \includegraphics[width=1\linewidth]{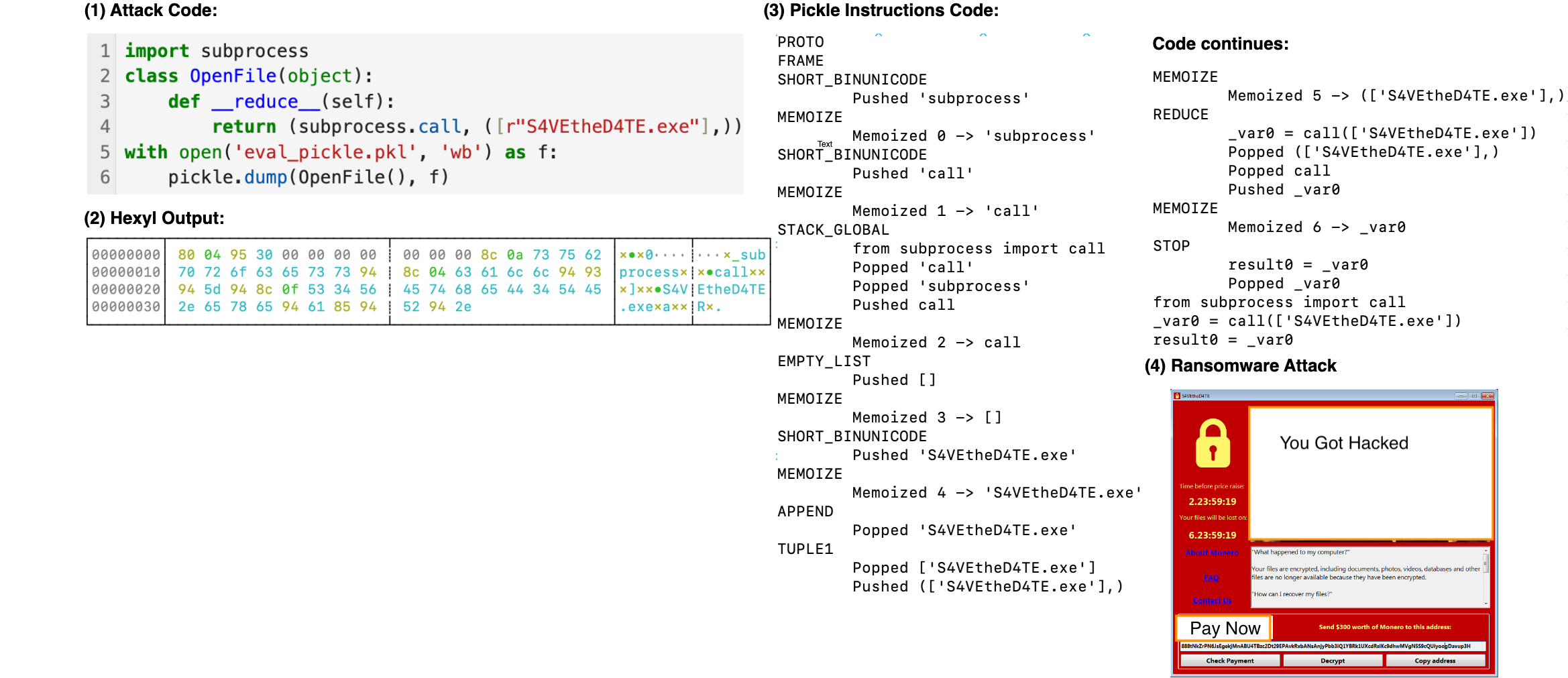}
    \caption{Visualization of Pickle model remote code execution. Step (1) illustrates how easy it is to create the attack, Step (2) is the hex view, Step (3) visualizes how Fickling prints the attack steps, and Step (4) is the execution of the ransomware.}
    \label{fig:pickle_ransom}
\end{figure*}

The Pickle module provides functions like \textbf{pickle.dump()} to serialize objects to a file-like object and \textbf{pickle.loads()} to deserialize objects from a byte stream. Pickle simplicity comes with a cost since the Pickle deserializer is called whenever the Pickle load function is invoked. By definition, serialization supports running arbitrary code instructions specified by the serializer (the process that runs the file). Therefore, running an attack is extremely simple and can be done in a few lines using the Python \_\_reduce\_\_ function as seen in Fig. \ref{fig:pickle_ransom} step one, lines three and four. However, overloading the Python \_\_reduce\_\_ function means that object can't be loaded since the Python \_\_reduce\_\_ function is responsible for defining the construction of the object. Once the model is loaded into memory, the subprocess call will execute the ransomware executable \cite{S4VEtheD4TE} as seen in Fig. \ref{fig:pickle_ransom} step four, and the returned value from pickle.load will be the result of the subporcess call. If we examine the raw data of the model using a hex viewer, we can observe that the subprocess call can be seen. However, not all data scientists may be familiar with the validation process and be aware of such details. Some software packages may automatically retrieve the model from a remote repository without notifying the user.

PyTorch uses Pickle internally to save modules \cite{pytorch_serialization}, tensors, and other PyTorch objects. The library defines the functions torch.save/torch.load for these purposes. A PyTorch saved model in the most up-to-date ZIP file format contains a "data.pkl" Pickle file, and some more files or metadata. This Pickle file, in particular, is the potential threat for MSA. 

The PM can be exploited using four
opcodes that allow the execution of arbitrary Python code outside of the PM and push the result onto the PM's stack. These opcodes are STACK\_GLOBAL, GLOBAL, GLOBAL\_OPCODE, and REDUCE. GLOBAL imports a Python module or class, while REDUCE applies a set of arguments to a callable previously imported through GLOBAL. It is worth noting that even if a Pickle file does not use the REDUCE opcode, importing a module alone can and will execute arbitrary code in that module, making GLOBAL alone dangerous. Using STACK\_GLOBAL, the pickled data includes references to global objects, allowing unpickling to work even if the module's namespace changes between pickling and unpickling. For instance, one can use GLOBAL to import the exec function from \_\_builtins\_\_" and then use REDUCE to call exec with an arbitrary string containing Python code to run. We can observe the \_\_reduce\_\_ function injected the malicious code as can be seen in Fig. \ref{fig:pickle_ransom} step three, where the STACK\_GLOBAL and REDUCE OPCODE are found. The full description of the other opcodes can be found here \cite{pickleopcode}.

\section{Architecture \& Solutions}
\label{Architecture}
The architecture section of our system comprises two security components that function collaboratively. The first component is a CDR \ref{CDR} flow designed to secure models that enter the organization or are downloaded onto a computer or server. The solution does not assume that the model files are benign, and it always distrusts them. The second component is the MTD \ref{MTD}, which provides comprehensive protection against physical attacks on the model weights and prevents unauthorized use of the trained model. It is used to save and load models.
The CDR component is also used to complement security when the MTD is not utilized, and the model's authenticity is unknown.

\subsection{Safe-Unpickling and CDR}
\label{CDR}

The proposed CDR solution is based on a secure Pickle deserialization process. It provides file security visibility, removes various possible attack vectors, and exports a secure Pickle or SafeTensors \cite{Safetensors} model. 

\begin{figure}
    \centering
    \includegraphics[width=\columnwidth]{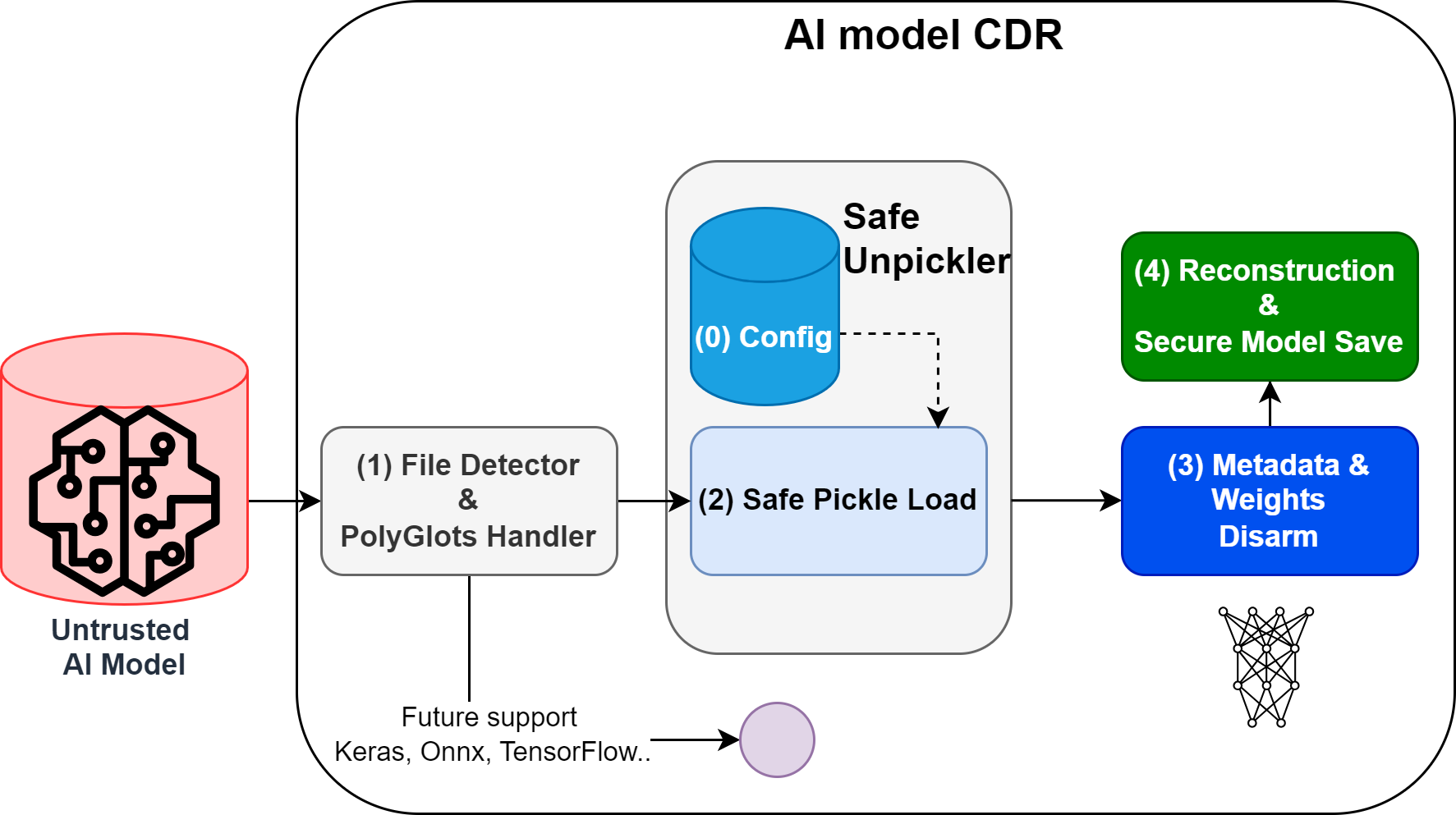}
    \caption{AI model CDR software package Architecture designed to eliminate serialization attacks.}
    \label{fig:cdr_arch}
\end{figure}

Fig. \ref{fig:cdr_arch} describes the process of loading and securing a model without knowing whether it is benign or malicious.

The process consists of several stages. 
In the initial \textbf{step (0)}, the user picks a policy for the safe-unpickler, which function/argument pairs to allow.
The first \textbf{stage (1)} focuses on detecting the model type. This work focuses on Pickle, but additional file formats will be handled in future work. 
Once the Pickle is detected, the second \textbf{stage (2)}
deserializes the model using a secure unpickling process that extends the original Pickle Unpickler class. The safe-unpickler overrides the Unpickler.load\_reduce class method. Doing so allows the authorization of a \_\_reduce\_\_ call based on the requested function and its arguments. Individually allowing \_\_reduce\_\_ calls gives greater control than the alternative approach of allowing/forbidding imports through the find\_class method as seen in other related works \cite{huang2022pain}. Moreover, it allows us to define specific behavior for cases where a \_\_reduce\_\_ call is blocked, which in some cases, enables partially loading Pickle files that contain some malicious sub-objects and some benign sub-objects. We compile sets of allowed/forbidden function calls for common use cases, e.g., PyTorch models.
See Table \ref{tab:CDR_modes} for example rulesets. We show a ruleset name, description, and blocked/allowed function calls. The Fickling ruleset is based on the Fickling unsafe calls analysis \cite{Fickling}, it recommends blocking builtins.eval, builtins.exec, os.system, and also specific library calls such as torch.load; we still want to maintain the main Python functionallity to allow constructors like builtins.list, which don't pose a threat. The TORCH ruleset specifically aims to allow using PyTorch, including loading serialized models, so we need to allow functions that torch.load uses such as torch.\_utils.\_rebuild\_tensor\_v2.

\begin{table*}[]
\caption{Example rulesets for allowed or forbidden functions and modules. Safe-Unpickler decides if a given function and argument tuple is allowed or blocked based on some chosen ruleset. Rulesets are defined by the user to match his scenario.}
\label{tab:CDR_modes}

\resizebox{\textwidth}{!}{%
\begin{tabular}{@{}llll@{}}
\toprule
Mode     & Description                                                                                                & \multicolumn{2}{l}{Example Rules}                                                                            \\
         &                                                                                                            & Block calls                                                            & Allow calls                         \\ \midrule
FICKLING & \begin{tabular}[c]{@{}l@{}}Ruleset derived from the\\ fickling analysis of potential threats.\end{tabular} & \begin{tabular}[c]{@{}l@{}}builtins.eval,\\ builtins.exec\end{tabular} & builtins.list                       \\
         &                                                                                                            & numpy.testing.\_private.utils.runstring                                & builtins.dict                       \\
         &                                                                                                            & os.system                                                              & numpy                               \\
TORCH    & \begin{tabular}[c]{@{}l@{}}Designed to allow loading\\ PyTorch models\end{tabular}                         & FICKLING Blocked calls                                                 & torch.\_utils.\_rebuild\_tensor\_v2 \\
         &                                                                                                            & ...                                                                    & collections.OrderedDict             \\ \bottomrule
\end{tabular}%
}

\end{table*}

\noindent In \textbf{stage six (3)}, hidden attacks on the model weights are disarmed based on previous CDR work focusing on disarming steganography attacks inside the model weights \cite{dubin2023disarmingAccess}. Note that this step is out of this work's scope and is not evaluated in this work, but a full CDR solution for AI models should enable them.

The last \textbf{step (4)} reconstructs the model weights as a new file. Finally, a clean model is exported. For added security, we offer a configuration option to save the model using Safetensors ~\cite{Safetensors} instead of Pickle. Safetensors is a serialization format designed to securely and efficiently handle tensor data. Unlike traditional formats such as Pickle, which can execute arbitrary code during deserialization, Safetensors is structured to prevent such security risks. This makes it particularly well-suited for use cases involving machine learning models and data, where the integrity and safety of the serialized data are of utmost importance. After the CDR, the model is evaluated using Fickling \cite{Fickling} to verify the secured re-created model.

\subsection{MTD}
\label{MTD}

\begin{figure}
    \centering
    \includegraphics[width=1.0\linewidth]{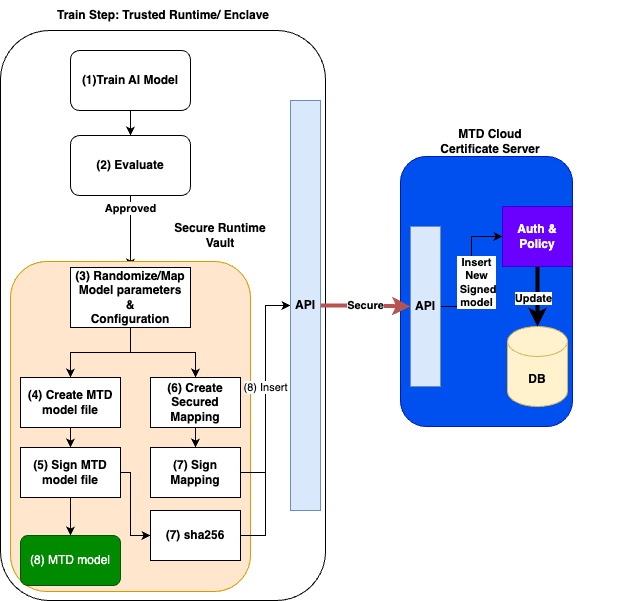}
    \caption{MTD model creation inside a secure enclave}
    \label{fig:mtd_creation}
\end{figure}

\begin{figure}
    \centering
    \includegraphics[width=1.0\linewidth]{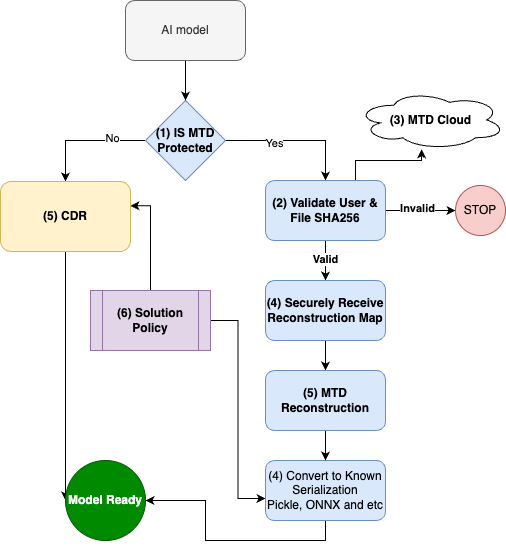}
    \caption{MTD model loading and fallback to CDR}
    \label{fig:mtd_load_and_cdr}
\end{figure}

\noindent The architecture is divided into two parts: MTD creation (Fig. \ref{fig:mtd_creation}) and MTD loading (Fig. \ref{fig:mtd_load_and_cdr}). If the model is not MTD protected, a fallback step uses CDR (Fig. \ref{fig:mtd_load_and_cdr}).

The model training process powered by MTD is demonstrated in Fig. \ref{fig:mtd_creation} and occurs within a secure runtime environment or enclave \cite{intelsgx}. An alternative option is to train the model first and then apply CDR against serialization attacks. Once the training is complete (1) and model evaluation passes (2), the MTD process (3) begins. The NSA \cite{NSA_AI_MODEL_sec} recommends separating the model and its weights. Therefore, the MTD algorithm randomizes the model weights, separates the model mapping from the randomized weights, and prevents unauthorized use of the trained model.
To provide additional security, the MTD model's weights can be encrypted.
This produces a file (4) that is protected by MTD. We sign and send this file to the cloud (5, 8) along with a mapping (6). We also sign and store the mapping in the cloud (6, 7).
To further protect the model, upon saving the MTD model, we save a sha256 hash of the model weights and verify the hash on load.

Figure \ref{fig:mtd_load_and_cdr} shows how the MTD-protected model is loaded. First, it checks (1) if the model is MTD protected. If it's not, the CDR (5) (Figure \ref{fig:cdr_arch}) is used instead. If the model is protected, it's validated (2) with the MTD cloud API (3). If any modifications are detected, an alert is sent, and the process is stopped. If the model is valid, the mapping is received, and the model is reconstructed. Once the model is reconstructed, it can be saved to a file or kept in memory. It's important to note that this process assumes that if the user is compromised, it's already too late, and the focus is on securing the model and not the device. AI inference can also run inside a secure enclave. Our future work will focus on securing AI PC, which is expected to be the future of running large language models by utilizing a Neural Processing Unit (NPU) (see \cite{aipc}).

\section{Dataset}
\label{Dataset}
\noindent For evaluation of our suggested methods, we create 3 datasets: benign AI models, malicious pickles, and AI models attacked with steganography.

\noindent \textbf{Dataset 1 - Benign PyTorch models}: A dataset of benign pre-trained PyTorch models. We compile this dataset using the torchvision and HuggingFace APIs. Table \ref{tab:ds_pt} contains a subset of the pre-trained model architectures we used. In total, this dataset contains 129 vision, text, and audio models. We use this to validate our methods on benign data and it is used as a base for creating a dataset of models attacked with steganography (dataset 3) for further evaluation.

\begin{table}[]
\caption{Subset of PyTorch model architectures used to validate the safe-unpickler-based CDR and MTD methods.}
\label{tab:ds_pt}
\centering
\resizebox{\columnwidth}{!}{%
\begin{tabular}{llll}
\toprule
Model Type                                                                                                  & Model Architecture & \#Models & Size {[}MB{]} \\ \midrule
\rowcolor[HTML]{EFEFEF} 
\cellcolor[HTML]{EFEFEF}                                                                                    & AlexNet \cite{alexnet}            & 1        & 230           \\
\rowcolor[HTML]{EFEFEF} 
\cellcolor[HTML]{EFEFEF}                                                                                    & ConvNext \cite{convnet}          & 4        & 110-750       \\
\rowcolor[HTML]{EFEFEF} 
\cellcolor[HTML]{EFEFEF}                                                                                    & DenseNet \cite{densenet}          & 4        & 30-110        \\
\rowcolor[HTML]{EFEFEF} 
\cellcolor[HTML]{EFEFEF}                                                                                    & EfficientNet \cite{efficientnet, efficientnet2}      & 12       & 20-450        \\
\rowcolor[HTML]{EFEFEF} 
\multirow{-5}{*}{\cellcolor[HTML]{EFEFEF}\begin{tabular}[c]{@{}l@{}}Vision\\ (Classification)\end{tabular}} & RegNet \cite{regnet}            & 34       & 20-2460       \\
\begin{tabular}[c]{@{}l@{}}Vision\\ (Segmentation)\end{tabular}                                             & DeepLabV3 \cite{deeplabv3}         & 3        & 40-230        \\
\rowcolor[HTML]{EFEFEF} 
\begin{tabular}[c]{@{}l@{}}Vision\\ (Detection)\end{tabular}                                                & Faster R-CNN \cite{faster_rcnn}      & 4        & 70-160        \\
\begin{tabular}[c]{@{}l@{}}Vision\\ (Video)\end{tabular}                                                    & SwinTransformer \cite{swin}   & 4        & 110-360       \\
\rowcolor[HTML]{EFEFEF} 
Text                                                                                                        & BERT \cite{bert}              & 8        & 440-1440      \\
Audio                                                                                                       & Wav2Vec \cite{wav2vec}           & 2        & 380-1270      \\ \bottomrule
\end{tabular}%
}
\end{table}

\begin{figure}
    \centering
    \includegraphics[width=1.0\linewidth]{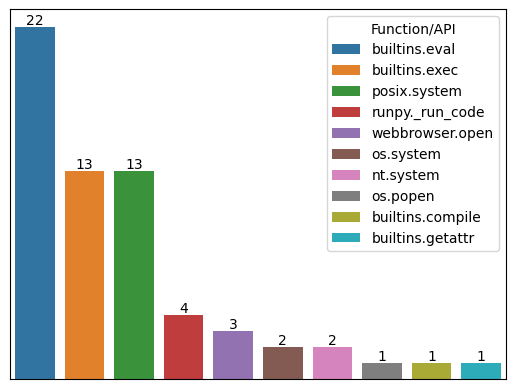}
    \caption{Distribution of function calls in MalHug \cite{malhug} malicious pickles.}
    \label{fig:malhug_dist}
\end{figure}

\noindent \textbf{Dataset 2.1 - MalHug Malicious Pickles}: HuggingFace contains various repositories with malicious pickles, some detected and some undetected. \cite{malhug} performed a systematic scraping of these pickles using a comprehensive static analysis. We specifically use the Pickle files (.pkl, .pickle) and PyTorch files (pytorch\_model.bin, .pt, .pth, .ckpt) they found. In total, we have 61 different malicious pickles that we use to validate our safe-unpickler-based CDR method. See Figure \ref{fig:malhug_dist} for a distribution of the function calls used in the malicious pickles from MalHug.

\noindent \textbf{Dataset 2.2 - zpbrent Malicious Pickles}: \cite{Pickle_serialization_zpbrent} presented clever Pickle serialization attacks. We use a subset of the malicious pickles available in the huggingface.co/zpbrent repository for evaluation of our safe-unpickler-based CDR method. In particular we use:
\begin{enumerate}
    \item zpbrent/RagReuse/psgs\_w100.tsv.pkl
    \item zpbrent/reuse/vocab.pkl
\end{enumerate}

\noindent \textbf{Dataset 3 - Attacked models for MTD method evaluation}: 
We use LSB model weight steganography attacks \cite{gilkarov2024steganalysis} from our previous work on the benign models from dataset 1 to validate the MTD verification, ensuring that the MTD process works as intended with untempered models and also that the method successfully detects changes (i.e., physical attacks) in the model as intended.

\section{Evaluation}
\label{Evaluation}
\noindent This section evaluates our proposed safe-unpickler-based CDR and AI model MTD methods. The experiments are designed to prove the methods have practical value for protecting AI model ecosystems from physical attacks.

\textbf{Experimental Setup}: All experiments are run on a Ubuntu Linux 22.04 server, equipped with an Intel Xeon Silver 4310 CPU (48 cores @ 2.10GHz), and 128 GB RAM.

\subsection{Safe-unpickler-based CDR}
The proposed safe-unpickler-based CDR is evaluated by comparing it to the traditional "find\_class" restricted-unpickler implementations seen so far. We use the restricted-unpicklers inside a safe sandbox Virtual Machine (VM) environment on the benign dataset 1 and malicious datasets 2.1 and 2.2 (See Section \ref{Dataset}). We carefully validate whether or not an attack happened by monitoring system resources, tracing the output channels, and running in a non-elevated environment without permissions, which throws exceptions upon execution of code that accesses the system components such as storage, internet, etc. Additionally, we establish that preventing the execution of unwanted function calls is purely based on policy. A policy that doesn't allow "builtins.eval" will successfully prevent malicious pickles that use this call. \cite{huang2022pain} asserted that restricted-unpicklers that check both "module" and "name" simultaneously are generally safe. Our method enforces this methodology, and Uranium unpickler (*) does so too.
Hence, the two main focuses of the experiment are: 
\begin{enumerate}
    \item Ensuring the method successfully loads benign pickles
    \item Finding out if our proposed safe-unpickler can successfully partially load objects in contrast with the previous methods. This benefit is crucial for CDR. 
\end{enumerate}
\subsubsection{Baseline restricted-unpicklers}
\label{sec:baseline_unpicklers}
For comparison, we use two "find\_class" restricted-unpicklers cited in \cite{huang2022pain}:
\begin{enumerate}
    \item Uranium \cite{uranium} - Indicated by \cite{huang2022pain} as the most robust of the compared restricted-unpicklers. Denoted in experiments as \textbf{(*)}.
    \item Fate \cite{fate} - designed for safely loading PyTorch models. Indicated by \cite{huang2022pain} as bypassable. Denoted in experiments as \textbf{(**)}.
\end{enumerate}

\subsubsection{Fickling Scan}
\label{sec:fickling_mapping}
We use Fickling \cite{Fickling} scans to quantify how malicious a Pickle file is before and after CDR.
In the following section, we map the fickling severity levels as follows:
\begin{itemize}
    \item -1: Pickle Deserialization Failed
    \item 0: LIKELY SAFE
    \item 1: POSSIBLY UNSAFE
    \item 2: SUSPICIOUS
    \item 3: LIKELY UNSAFE
    \item 4: LIKELY OVERTLY MALICIOUS
    \item 5: OVERTLY MALICIOUS
\end{itemize}
We note that benign PyTorch models get a fickling severity level of 3. Therefore, fickling analysis is not a perfect indicator of malicious activity, but it is a useful heuristic tool. 
\subsubsection{Experimental Evaluation}

Initially, we tested our proposed safe-unpickler-based CDR on the benign dataset (dataset 1), and indeed, the models are loaded correctly with a 100\% success rate using the PyTorch policy we discussed in Section \ref{CDR}. The baseline restricted-unpicklers also successfully load the models using the same policy.
We want to see whether models get a lower "fickling malicious severity" score after CDR. See Table \ref{tab:exp_cdr_rates}, we can see our method successfully loaded 9/26 and 10/35 pickles and PyTorch models, respectively, in contrast with the "find\_class" restricted-unpicklers that can't partially load a contaminated Pickle.
Moreover, looking at Table \ref{tab:exp_cdr}, we can see that our suggested methods mainly succeed in loading malicious pickles with fickling severity levels 4 and lower and that the baseline "find\_class" restricted-unpicklers (see Section \ref{sec:baseline_unpicklers}) failed to load any Pickle as expected. Successfully loaded PyTorch models after CDR get a fickling severity of level 3, which is the level given to regular benign PyTorch models.
We move on to look at our specific datasets. Looking at Table \ref{tab:pickle_cdr_zpbrent}, our methods successfully blocked the zpbrent malicious pickles (dataset 2.2), and also successfully loaded them. The pickles got a fickling score of 4 before CDR and 3 after CDR, which suggests the CDR process successfully lowered the threat.

\begin{table}[h]
\caption{Fickling scan results before and after deserialization of MalHug malicious pickles (dataset 2.1) using our proposed safe-unpickler and related methods. See section \ref{sec:fickling_mapping} for the fickling error code mapping.}
\label{tab:exp_cdr}
\centering
\begin{tabular}{@{}lllllll@{}}
\toprule
\multirow{2}{*}{\begin{tabular}[c]{@{}l@{}}Fickling\\ (before $\rightarrow$ after)\end{tabular}} & \multicolumn{3}{l}{\#Pickles} & \multicolumn{3}{l}{\begin{tabular}[c]{@{}l@{}}\#PyTorch Models\end{tabular}} \\ \cmidrule(l){2-7}
                                                                                                 & Ours    & (*)   & (**)   & Ours                    & (*)                    & (**)                   \\ \midrule 
5 $\rightarrow$ 3                                                                                & 0       & 0       & 0         & \textbf{2}                       & 0                        & 0                         \\
5 $\rightarrow$ -1                                                                               & 11      & 11      & 11        & 22                      & 24                       & 24                        \\
4 $\rightarrow$ 3                                                                                & \textbf{8}       & 0       & 0         & \textbf{1}                       & 0                        & 0                         \\
4 $\rightarrow$ -1                                                                               & 2       & 10      & 10        & 3                       & 4                        & 4                         \\
3 $\rightarrow$ 3                                                                                & \textbf{1}       & 0       & 0         & \textbf{7}                       & 0                        & 0                         \\
3 $\rightarrow$ -1                                                                               & 4       & 5       & 5         & 0                       & 7                        & 7                         \\ \bottomrule
\end{tabular}%
\end{table}

\begin{table}[]
\caption{Safe-unpickler attack prevention rate and successful Pickle load rate on dataset 2.1 using our proposed method and related methods.}
\label{tab:exp_cdr_rates}

\centering
\resizebox{\columnwidth}{!}{%
\begin{tabular}{@{}llll@{}}
\toprule
Unpickler   & Attacks Mitigated & \multicolumn{2}{l}{Successful Load} \\ \cmidrule(l){3-4} 
            &                   & \#Pickles        & PyTorch          \\ \midrule
Ours        & 100\%             & \textbf{9/26}    & \textbf{10/35}   \\
Uranium (*) & 100\%             & 0/26             & 0/35             \\
Fate (**)   & 100\%             & 0/26             & 0/35             \\ \bottomrule
\end{tabular}%
}
\end{table}

\begin{table}[]
\caption{Pickle CDR Results on zpbrent malicious pickles (dataset 2.2) using our proposed Safe-Unpickler.}
\label{tab:pickle_cdr_zpbrent}
\centering
\resizebox{\columnwidth}{!}{%
\begin{tabular}{@{}lcl@{}}
\toprule
Hugging Face URI                    & \multicolumn{1}{l}{Attack Blocked?} & \begin{tabular}[c]{@{}l@{}}Fickling\\ (before $\rightarrow$ after)\end{tabular} \\ \midrule
zpbrent/RagReuse/psgs\_w100.tsv.pkl & Yes                                 & 4 $\rightarrow$ 3                                                               \\
zpbrent/reuse/vocab.pkl             & Yes                                 & 4 $\rightarrow$ 3                                                               \\ \bottomrule
\end{tabular}%
}
\end{table}

\subsection{MTD}
We evaluate the MTD method for protecting training models with 2 main concerns: validity and runtime. Validity means we assert that the method prevents the attacks it's supposed to prevent and that the models stay intact after the whole process. We measure runtime to analyze the time overhead that the MTD methods cause.
We created automated Quality Assurance (QA) tests to check the method's validity.
The tests check the following:
\begin{enumerate}
    \item MTD model is the same after obfuscation and deobfuscation.
    \item MTD model is the same after saving and loading.
    \item Attack simulation: An MTD Model is constructed, obfuscated, and saved. Then the model weights are attacked, and we assert that the change is detected upon load.
\end{enumerate}

\begin{figure}
    \centering
    \includegraphics[width=1\linewidth]{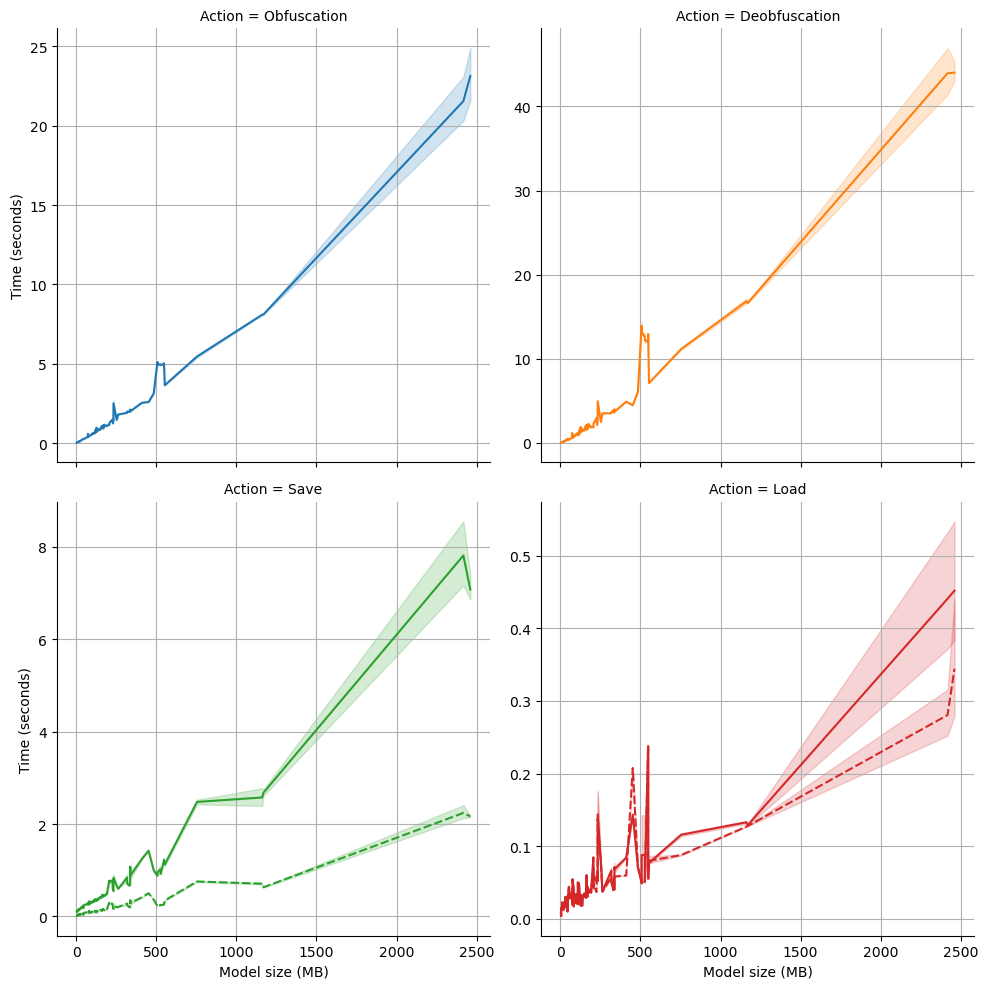}
    \caption{Mean time (seconds) measurement vs. model size (MB) of different MTD actions: obfuscation, deobfuscation, load, and save. We show the MTD variant (solid line) and the regular variant (dashed line) for load and save. The action procedures were repeated 10 times with each model in the MTD datasets, and we show confidence intervals. We can see a linear increase in runtime as the model size grows.}
    \label{fig:mtd_time}
\end{figure}

For steps 1 and 2, we use the benign PyTorch models dataset (dataset 1, see Section \ref{Dataset}); for step 3, we use dataset 3.
All tests passed. Figure \ref{fig:mtd_time} plots the mean runtime of the different MTD functions on all models in dataset 1, repeated 10 times. All IO operations, i.e., saving and loading, were done with a virtual file since we are only concerned with the runtime overhead our method adds. We can see obfuscation and deobfuscation take less than a minute for models of size up to 2.5GB, we feel this is reasonable, for reference, LLaMa3.2 \cite{llama3modelcard} with 3B parameters is 2GB in size. Moreover, looking at loading, the MTD procedure added an insignificant performance overhead, and saving added about a 2-fold performance overhead.

\section{Limitations}
\label{Limitations}
The proposed solution introduces a novel architecture for physically securing AI model files. The architecture protects against serialization attacks by utilizing CDR and preventing all file alterations when the model is secured with MTD and the proposed architecture. This builds upon previous work on CDR against steganography attacks \cite{gilkarov2024steganalysis} and steganalysis \cite{gilkarov2024steganalysis} (steganography attack detection). We can protect against all known file-born physical attacks by using the solutions together. Since AI model file security is still in its early stages, we expect to see more sophisticated attacks and vulnerabilities in this domain.
The proposed safe-unpickler method requires ongoing maintenance of evolving configuration, but with this approach, this may be necessary when new attacks are found. The proposed MTD architecture requires the user to fully use the various AI model functions such as training, and serialization within it, but widely-used methods like SafeTensors \cite{Safetensors} do so too. Moreover, both proposed methods add a certain performance overhead that is added to all models since they work in a zero-trust manner. However, the performance overhead is small, as described in the result section.
Additionally, this paper focuses on Pickle and PyTorch, the most commonly exploited file formats. Although other formats, such as TensorFlow \cite{tensorflowfileformat}, Keras (HDf5) \cite{hdf5}, and ONNX \cite{onnx}, are known to be exploited, the risk is significantly lower. Our future work will extend and support them. It is important to note that this work focuses solely on the security of the AI model's physical file. Other adversarial attacks exist, such as backdoors in machine learning models \cite{wenger2021backdoor} and dataset poisoning attacks \cite{zheng2022concealed}. Dataset attacks are outside this paper's scope, but future work should view them as part of a holistic platform.

\section{Conclusions}
\label{Conclusions}
Our work proposes a new solution for securing AI model files by preventing serialization attacks in Pickle and PyTorch using CDR and preventing file modifications through the use of MTD and authentication. This architecture aims to prevent supply chain attacks and secure AI model file transfers from serialization attacks that threaten the industry. Despite the existence of safer file serialization methods, Pickle is still widely used. Our proposed architecture, which includes MTD and CDR, provides a significant step towards AI model file security since users typically download files from the network, and the Pickle file is known to be vulnerable. The CDR in our work prevents serialization attacks that allow the execution of malicious malware while the model is loading.

By incorporating MTD, we follow the NSA's \cite{NSA_AI_MODEL_sec} recommendation to separate the model architecture and its weights. This is accomplished by randomizing the model weights and separating the model from the deconstruction mapping. This ensures that only authenticated users can reconstruct and use the model when it is distributed. The architecture provides a 100\% prevention rate.

Our methods were evaluated on a wide range of data which includes diverse benign models, steganography attacks, and Pickle, PyTorch serialization attacks from the literature and real-world scenarios found in HuggingFace \cite{Huggingfacecite}

\section{Acknowledgement} This work is under US Provisional Patent Application No. 63/536,420

\bibliographystyle{IEEEtran}
\bibliography{main}

\begin{thebibliography}{10}
\providecommand{\url}[1]{#1}
\csname url@samestyle\endcsname
\providecommand{\newblock}{\relax}
\providecommand{\bibinfo}[2]{#2}
\providecommand{\BIBentrySTDinterwordspacing}{\spaceskip=0pt\relax}
\providecommand{\BIBentryALTinterwordstretchfactor}{4}
\providecommand{\BIBentryALTinterwordspacing}{\spaceskip=\fontdimen2\font plus
\BIBentryALTinterwordstretchfactor\fontdimen3\font minus \fontdimen4\font\relax}
\providecommand{\BIBforeignlanguage}[2]{{%
\expandafter\ifx\csname l@#1\endcsname\relax
\typeout{** WARNING: IEEEtran.bst: No hyphenation pattern has been}%
\typeout{** loaded for the language `#1'. Using the pattern for}%
\typeout{** the default language instead.}%
\else
\language=\csname l@#1\endcsname
\fi
#2}}
\providecommand{\BIBdecl}{\relax}
\BIBdecl

\bibitem{chen2024zddr}
M.~Chen, G.~He, and J.~Wu, ``Zddr: A zero-shot defender for adversarial samples detection and restoration,'' \emph{IEEE Access}, 2024.

\bibitem{nguyen2023physical}
K.~Nguyen, T.~Fernando, C.~Fookes, and S.~Sridharan, ``Physical adversarial attacks for surveillance: A survey,'' \emph{IEEE Transactions on Neural Networks and Learning Systems}, 2023.

\bibitem{yao2024survey}
Y.~Yao, J.~Duan, K.~Xu, Y.~Cai, Z.~Sun, and Y.~Zhang, ``A survey on large language model (llm) security and privacy: The good, the bad, and the ugly,'' \emph{High-Confidence Computing}, p. 100211, 2024.

\bibitem{NSA_AI_MODEL_sec}
\BIBentryALTinterwordspacing
N.~S. Agency, ``Deploying ai systems securely, best practices for deploying secure and resilient ai systems,'' 2024, accessed: 2024-05-01. [Online]. Available: \url{https://media.defense.gov/2024/Apr/15/2003439257/-1/-1/0/CSI-DEPLOYING-AI-SYSTEMS-SECURELY.PDF}
\BIBentrySTDinterwordspacing

\bibitem{mitre_atlas}
\BIBentryALTinterwordspacing
M.~ATLAS, ``Mitre atlas,'' 2024, accessed: 2024-05-01. [Online]. Available: \url{https://atlas.mitre.org/}
\BIBentrySTDinterwordspacing

\bibitem{hiddenlayer}
\BIBentryALTinterwordspacing
{E. Wickens}, {M. Janus}, and {T. Bonner}, ``Weaponizing machine learning models with ransomware,'' 2022, accessed: 2024-05-01. [Online]. Available: \url{https://hiddenlayer.com/research/weaponizing-machine-learning-models-with-ransomware/}
\BIBentrySTDinterwordspacing

\bibitem{models_are_code}
\BIBentryALTinterwordspacing
T.~Bonner, ``Models are code,'' 2024, accessed: 2024-05-01. [Online]. Available: \url{https://hiddenlayer.com/research/models-are-code/}
\BIBentrySTDinterwordspacing

\bibitem{Pickle_serialization_vul}
\BIBentryALTinterwordspacing
M.~Slaviero, ``Sour pickles, a serialized exploitation guide in one part,'' accessed: 2023-05-01. [Online]. Available: \url{https://media.blackhat.com/bh-us-11/Slaviero/BH_US_11_Slaviero_Sour_Pickles_Slides.pdf}
\BIBentrySTDinterwordspacing

\bibitem{pytorchhub}
\BIBentryALTinterwordspacing
PyTorch, ``Pytorch model sharing hub,'' 2022, accessed: 2022-01-15. [Online]. Available: \url{https://pytorch.org/hub/}
\BIBentrySTDinterwordspacing

\bibitem{malhug}
\BIBentryALTinterwordspacing
J.~Zhao, S.~Wang, Y.~Zhao, X.~Hou, K.~Wang, P.~Gao, Y.~Zhang, C.~Wei, and H.~Wang, ``Models are codes: Towards measuring malicious code poisoning attacks on pre-trained model hubs,'' in \emph{Proceedings of the 39th IEEE/ACM International Conference on Automated Software Engineering}, ser. ASE ’24.\hskip 1em plus 0.5em minus 0.4em\relax ACM, Oct. 2024, p. 2087–2098. [Online]. Available: \url{http://dx.doi.org/10.1145/3691620.3695271}
\BIBentrySTDinterwordspacing

\bibitem{Huggingfacecite}
\BIBentryALTinterwordspacing
H.~Face, ``Hugging face model zoo,'' accessed: 2023-05-01. [Online]. Available: \url{https://huggingface.co/models}
\BIBentrySTDinterwordspacing

\bibitem{hftopmodel}
\BIBentryALTinterwordspacing
J.~Grosman, ``Fine-tuned xlsr-53 large model for speech recognition in english,'' 2023, accessed: 2023-10-15. [Online]. Available: \url{https://huggingface.co/models?sort=downloads}
\BIBentrySTDinterwordspacing

\bibitem{pickle_standard}
\BIBentryALTinterwordspacing
Pickle, ``Pickle- python object serialization,'' accessed: 2023-05-01. [Online]. Available: \url{https://docs.python.org/3/library/pickle.html}
\BIBentrySTDinterwordspacing

\bibitem{Huggingfacepickle}
\BIBentryALTinterwordspacing
HuggingFace, ``{Fickling pickle scanning},'' 2022, accessed: 2022-12-19. [Online]. Available: \url{https://huggingface.co/}
\BIBentrySTDinterwordspacing

\bibitem{ClamAV}
\BIBentryALTinterwordspacing
ClamAV, ``{ClamAV® open-source antivirus engine for detecting trojans, viruses, malware \& other malicious threats.}'' 2022, accessed: 2022-12-19. [Online]. Available: \url{https://www.clamav.net/}
\BIBentrySTDinterwordspacing

\bibitem{pickle-attack}
\BIBentryALTinterwordspacing
E.~Sultanik, ``{Never a dill moment: Exploiting machine learning pickle files},'' 2022, accessed: 2022-12-19. [Online]. Available: \url{https://blog.trailofbits.com/2021/03/15/never-a-dill-moment-exploiting-machine-learning-pickle-files/}
\BIBentrySTDinterwordspacing

\bibitem{jfrog_ai_model}
\BIBentryALTinterwordspacing
D.~Cohen, ``Data scientists targeted by malicious hugging face ml models with silent backdoor,'' 2024, accessed: 2024-05-01. [Online]. Available: \url{https://jfrog.com/blog/data-scientists-targeted-by-malicious-hugging-face-ml-models-with-silent-backdoor/}
\BIBentrySTDinterwordspacing

\bibitem{OWSAPtop10ml}
\BIBentryALTinterwordspacing
OWSAP, ``Owasp machine learning security top ten,'' 2023, accessed: 2023-10-15. [Online]. Available: \url{https://owasp.org/www-project-machine-learning-security-top-10/}
\BIBentrySTDinterwordspacing

\bibitem{ATLAS}
\BIBentryALTinterwordspacing
MITRE, ``Mitre atlas,'' 2023, accessed: 2023-10-15. [Online]. Available: \url{https://atlas.mitre.org/}
\BIBentrySTDinterwordspacing

\bibitem{mitre_atlas_unsafeml}
\BIBentryALTinterwordspacing
M.~ATLAS, ``Mitre atlas, user execution: Unsafe ml artifacts,'' 2024, accessed: 2024-05-01. [Online]. Available: \url{https://atlas.mitre.org/techniques/AML.T0011.000}
\BIBentrySTDinterwordspacing

\bibitem{dubin2023disarmingAccess}
R.~Dubin, ``Disarming attacks inside neural network models,'' \emph{IEEE Access}, 2023.

\bibitem{PengZhou}
\BIBentryALTinterwordspacing
P.~Zhou, ``How to make hugging face to hug worms: Discovering and exploiting unsafe pickle.loads over pre-trained large model hubs,'' accessed: 2024-08-01. [Online]. Available: \url{https://i.blackhat.com/Asia-24/Presentations/Asia-24-Zhou-HowtoMakeHuggingFace.pdf}
\BIBentrySTDinterwordspacing

\bibitem{PickleRestriction}
\BIBentryALTinterwordspacing
Pickle, ``Restricting globals,'' accessed: 2024-09-15. [Online]. Available: \url{https://docs.python.org/3/library/pickle.html#restricting-globals}
\BIBentrySTDinterwordspacing

\bibitem{Pickle_attack_wild}
\BIBentryALTinterwordspacing
T.~B. Eoin~Wickens, Marta~Janus, ``Pickle files: the new ml model attack vector,'' accessed: 2023-05-01. [Online]. Available: \url{https://hiddenlayer.com/research/pickle-strike/}
\BIBentrySTDinterwordspacing

\bibitem{huang2022pain}
N.-J. Huang, C.-J. Huang, and S.-K. Huang, ``Pain pickle: Bypassing python restricted unpickler for automatic exploit generation,'' in \emph{2022 IEEE 22nd International Conference on Software Quality, Reliability and Security (QRS)}.\hskip 1em plus 0.5em minus 0.4em\relax IEEE, 2022, pp. 1079--1090.

\bibitem{Pikora}
\BIBentryALTinterwordspacing
splitline, ``Pikora a small compiler that can convert python scripts to pickle bytecode.'' accessed: 2024-09-15. [Online]. Available: \url{https://github.com/splitline/Pickora}
\BIBentrySTDinterwordspacing

\bibitem{petastorm}
\BIBentryALTinterwordspacing
Uber, ``Uber petastorm,'' accessed: 2024-09-15. [Online]. Available: \url{https://github.com/uber/petastorm/issues/741}
\BIBentrySTDinterwordspacing

\bibitem{Pickle-Fuzz}
\BIBentryALTinterwordspacing
Pickle-Fuzz, ``Pickle-fuzz,'' accessed: 2024-09-15. [Online]. Available: \url{https://github.com/moreati/pickle-fuzz}
\BIBentrySTDinterwordspacing

\bibitem{SleepyPickle}
\BIBentryALTinterwordspacing
B.~Milanov, ``Exploiting ml models with pickle file attacks: Part 1,'' accessed: 2024-09-15. [Online]. Available: \url{https://blog.trailofbits.com/2024/06/11/exploiting-ml-models-with-pickle-file-attacks-part-1/}
\BIBentrySTDinterwordspacing

\bibitem{Fickling}
\BIBentryALTinterwordspacing
E.~Sultanik, ``{Fickling pickle scanning},'' 2021, accessed: 2022-12-19. [Online]. Available: \url{https://github.com/trailofbits/fickling}
\BIBentrySTDinterwordspacing

\bibitem{modelscan}
\BIBentryALTinterwordspacing
P.~AI, ``Modelscan: Protection against model serialization attacks,'' accessed: 2023-05-01. [Online]. Available: \url{https://github.com/protectai/modelscan/}
\BIBentrySTDinterwordspacing

\bibitem{fate}
\BIBentryALTinterwordspacing
FederatedAI, ``Fate, an industrial grade federated learning framework.'' accessed: 2024-12-11. [Online]. Available: \url{https://github.com/FederatedAI/FATE}
\BIBentrySTDinterwordspacing

\bibitem{uranium}
\BIBentryALTinterwordspacing
Ultimaker, ``Uranium, a python framework for building desktop applications.'' accessed: 2024-12-11. [Online]. Available: \url{https://github.com/Ultimaker/Uranium}
\BIBentrySTDinterwordspacing

\bibitem{gauravfuture}
\BIBentryALTinterwordspacing
A.~Gaurav, ``The future of network security: Why zero trust is becoming the new standard,'' accessed: 2024-10-01. [Online]. Available: \url{https://insights2techinfo.com/the-future-of-network-security-why-zero-trust-is-becoming-the-new-standard/}
\BIBentrySTDinterwordspacing

\bibitem{belkind2023open}
E.~Belkind, R.~Dubin, and A.~Dvir, ``Open image content disarm and reconstruction,'' \emph{arXiv preprint arXiv:2307.14057}, 2023.

\bibitem{dubin2023contentpdf}
R.~Dubin, ``Content disarm and reconstruction of pdf files,'' \emph{Ieee Access}, 2023.

\bibitem{dubin2024contentole}
------, ``Content disarm and reconstruction of microsoft office ole files,'' \emph{Computers \& Security}, vol. 137, p. 103647, 2024.

\bibitem{dubin2023contentrtf}
------, ``Content disarm and reconstruction of rtf files a zero file trust methodology,'' \emph{IEEE Transactions on Information Forensics and Security}, vol.~18, pp. 1461--1472, 2023.

\bibitem{DocBleach}
\BIBentryALTinterwordspacing
DocBleach, ``Docbleach open source cdr,'' accessed: 2023-05-01. [Online]. Available: \url{https://github.com/docbleach/DocBleach}
\BIBentrySTDinterwordspacing

\bibitem{Disinfect:pdfCDR}
\BIBentryALTinterwordspacing
Disinfect, ``open source pdf cdr tool,'' accessed: 2023-05-01. [Online]. Available: \url{https://github.com/gxjit/Disinfect}
\BIBentrySTDinterwordspacing

\bibitem{greshake2023not}
K.~Greshake, S.~Abdelnabi, S.~Mishra, C.~Endres, T.~Holz, and M.~Fritz, ``Not what you've signed up for: Compromising real-world llm-integrated applications with indirect prompt injection,'' in \emph{Proceedings of the 16th ACM Workshop on Artificial Intelligence and Security}, 2023, pp. 79--90.

\bibitem{kotek2023gender}
H.~Kotek, R.~Dockum, and D.~Sun, ``Gender bias and stereotypes in large language models,'' in \emph{Proceedings of The ACM Collective Intelligence Conference}, 2023, pp. 12--24.

\bibitem{cho2020toward}
J.-H. Cho, D.~P. Sharma, H.~Alavizadeh, S.~Yoon, N.~Ben-Asher, T.~J. Moore, D.~S. Kim, H.~Lim, and F.~F. Nelson, ``Toward proactive, adaptive defense: A survey on moving target defense,'' \emph{IEEE Communications Surveys \& Tutorials}, vol.~22, no.~1, pp. 709--745, 2020.

\bibitem{heydari2018moving}
V.~Heydari, ``Moving target defense for securing scada communications,'' \emph{IEEE Access}, vol.~6, pp. 33\,329--33\,343, 2018.

\bibitem{azab2016migrate}
M.~Azab and M.~Eltoweissy, ``Migrate: Towards a lightweight moving-target defense against cloud side-channels,'' in \emph{2016 IEEE security and privacy workshops (SPW)}.\hskip 1em plus 0.5em minus 0.4em\relax IEEE, 2016, pp. 96--103.

\bibitem{styugin2016new}
M.~Styugin, V.~Zolotarev, A.~Prokhorov, and R.~Gorbil, ``New approach to software code diversification in interpreted languages based on the moving target technology,'' in \emph{2016 IEEE 10th International Conference on Application of Information and Communication Technologies (AICT)}.\hskip 1em plus 0.5em minus 0.4em\relax IEEE, 2016, pp. 1--5.

\bibitem{banescu2018tutorial}
S.~Banescu and A.~Pretschner, ``A tutorial on software obfuscation,'' \emph{Advances in Computers}, vol. 108, pp. 283--353, 2018.

\bibitem{mordehai2017method}
G.~Mordehai, Y.~Elovici, and G.~Kedma, ``Method and system for protecting computerized systems from malicious code,'' Jul.~11 2017, uS Patent 9,703,954.

\bibitem{evans2011effectiveness}
D.~Evans, A.~Nguyen-Tuong, and J.~Knight, ``Effectiveness of moving target defenses,'' \emph{Moving Target Defense: Creating Asymmetric Uncertainty for Cyber Threats}, pp. 29--48, 2011.

\bibitem{protobuf}
\BIBentryALTinterwordspacing
Google, ``Protobuf,'' accessed: 2023-05-01. [Online]. Available: \url{https://protobuf.dev/}
\BIBentrySTDinterwordspacing

\bibitem{S4VEtheD4TE}
\BIBentryALTinterwordspacing
ANONYMOUSLGD, ``S4vethed4te ransomware written in c\# using windows forms.'' accessed: 2023-05-01. [Online]. Available: \url{https://github.com/ANONYMOUSLGD/S4VEtheD4TE}
\BIBentrySTDinterwordspacing

\bibitem{pytorch_serialization}
\BIBentryALTinterwordspacing
PyTorch, ``Pytorch saving/loading,'' 2024. [Online]. Available: \url{https://pytorch.org/docs/stable/notes/serialization.html#saving-and-loading-torch-nn-modules}
\BIBentrySTDinterwordspacing

\bibitem{pickleopcode}
\BIBentryALTinterwordspacing
Kaitai\_Project, ``Pickle opcodes,'' accessed: 2023-05-01. [Online]. Available: \url{https://formats.kaitai.io/python_pickle/}
\BIBentrySTDinterwordspacing

\bibitem{Safetensors}
\BIBentryALTinterwordspacing
H.~Face, ``Safetensors,'' accessed: 2024-08-01. [Online]. Available: \url{https://huggingface.co/docs/safetensors/en/index}
\BIBentrySTDinterwordspacing

\bibitem{intelsgx}
\BIBentryALTinterwordspacing
Intel, ``Reference architecture for privacy preserving machine learning with intel® sgx and tensorflow* serving,'' accessed: 2023-05-01. [Online]. Available: \url{https://www.intel.com/content/www/us/en/developer/articles/technical/privacy-preserving-ml-with-sgx-and-tensorflow.html}
\BIBentrySTDinterwordspacing

\bibitem{aipc}
\BIBentryALTinterwordspacing
``The ai pc powered by intel is here. now, ai is for everyone.'' accessed: 2023-05-01. [Online]. Available: \url{https://www.intel.com/content/www/us/en/products/docs/processors/core-ultra/ai-pc.html}
\BIBentrySTDinterwordspacing

\bibitem{alexnet}
\BIBentryALTinterwordspacing
A.~Krizhevsky, I.~Sutskever, and G.~E. Hinton, ``Imagenet classification with deep convolutional neural networks,'' in \emph{Advances in Neural Information Processing Systems}, F.~Pereira, C.~Burges, L.~Bottou, and K.~Weinberger, Eds., vol.~25.\hskip 1em plus 0.5em minus 0.4em\relax Curran Associates, Inc., 2012. [Online]. Available: \url{https://proceedings.neurips.cc/paper_files/paper/2012/file/c399862d3b9d6b76c8436e924a68c45b-Paper.pdf}
\BIBentrySTDinterwordspacing

\bibitem{convnet}
\BIBentryALTinterwordspacing
Z.~Liu, H.~Mao, C.-Y. Wu, C.~Feichtenhofer, T.~Darrell, and S.~Xie, ``A convnet for the 2020s,'' 2022. [Online]. Available: \url{https://arxiv.org/abs/2201.03545}
\BIBentrySTDinterwordspacing

\bibitem{densenet}
\BIBentryALTinterwordspacing
G.~Huang, Z.~Liu, L.~van~der Maaten, and K.~Q. Weinberger, ``Densely connected convolutional networks,'' 2018. [Online]. Available: \url{https://arxiv.org/abs/1608.06993}
\BIBentrySTDinterwordspacing

\bibitem{efficientnet}
\BIBentryALTinterwordspacing
M.~Tan and Q.~V. Le, ``Efficientnet: Rethinking model scaling for convolutional neural networks,'' 2020. [Online]. Available: \url{https://arxiv.org/abs/1905.11946}
\BIBentrySTDinterwordspacing

\bibitem{efficientnet2}
\BIBentryALTinterwordspacing
------, ``Efficientnetv2: Smaller models and faster training,'' 2021. [Online]. Available: \url{https://arxiv.org/abs/2104.00298}
\BIBentrySTDinterwordspacing

\bibitem{regnet}
\BIBentryALTinterwordspacing
I.~Radosavovic, R.~P. Kosaraju, R.~Girshick, K.~He, and P.~Dollár, ``Designing network design spaces,'' 2020. [Online]. Available: \url{https://arxiv.org/abs/2003.13678}
\BIBentrySTDinterwordspacing

\bibitem{deeplabv3}
\BIBentryALTinterwordspacing
L.-C. Chen, G.~Papandreou, F.~Schroff, and H.~Adam, ``Rethinking atrous convolution for semantic image segmentation,'' 2017. [Online]. Available: \url{https://arxiv.org/abs/1706.05587}
\BIBentrySTDinterwordspacing

\bibitem{faster_rcnn}
\BIBentryALTinterwordspacing
S.~Ren, K.~He, R.~Girshick, and J.~Sun, ``Faster r-cnn: Towards real-time object detection with region proposal networks,'' 2016. [Online]. Available: \url{https://arxiv.org/abs/1506.01497}
\BIBentrySTDinterwordspacing

\bibitem{swin}
\BIBentryALTinterwordspacing
Z.~Liu, J.~Ning, Y.~Cao, Y.~Wei, Z.~Zhang, S.~Lin, and H.~Hu, ``Video swin transformer,'' 2021. [Online]. Available: \url{https://arxiv.org/abs/2106.13230}
\BIBentrySTDinterwordspacing

\bibitem{bert}
\BIBentryALTinterwordspacing
J.~Devlin, M.-W. Chang, K.~Lee, and K.~Toutanova, ``Bert: Pre-training of deep bidirectional transformers for language understanding,'' 2019. [Online]. Available: \url{https://arxiv.org/abs/1810.04805}
\BIBentrySTDinterwordspacing

\bibitem{wav2vec}
\BIBentryALTinterwordspacing
A.~Baevski, H.~Zhou, A.~Mohamed, and M.~Auli, ``wav2vec 2.0: A framework for self-supervised learning of speech representations,'' 2020. [Online]. Available: \url{https://arxiv.org/abs/2006.11477}
\BIBentrySTDinterwordspacing

\bibitem{Pickle_serialization_zpbrent}
\BIBentryALTinterwordspacing
P.~Zhou, ``How to make hugging face to hug worms,'' accessed: 2024-12-06. [Online]. Available: \url{https://i.blackhat.com/Asia-24/Presentations/Asia-24-Zhou-HowtoMakeHuggingFace.pdf}
\BIBentrySTDinterwordspacing

\bibitem{gilkarov2024steganalysis}
D.~Gilkarov and R.~Dubin, ``Steganalysis of ai models lsb attacks,'' \emph{IEEE Transactions on Information Forensics and Security}, 2024.

\bibitem{llama3modelcard}
\BIBentryALTinterwordspacing
AI@Meta, ``Llama 3 model card,'' 2024. [Online]. Available: \url{https://github.com/meta-llama/llama3/blob/main/MODEL_CARD.md}
\BIBentrySTDinterwordspacing

\bibitem{tensorflowfileformat}
\BIBentryALTinterwordspacing
TensorFlow, ``Tensorflow model file formats,'' accessed: 2023-05-01. [Online]. Available: \url{https://www.tensorflow.org/hub/model_formats}
\BIBentrySTDinterwordspacing

\bibitem{hdf5}
\BIBentryALTinterwordspacing
HDF5, ``High-performance data management and storage suite,'' accessed: 2023-05-01. [Online]. Available: \url{https://portal.hdfgroup.org/documentation/}
\BIBentrySTDinterwordspacing

\bibitem{onnx}
\BIBentryALTinterwordspacing
ONNX, ``Open neural network exchange,'' accessed: 2023-05-01. [Online]. Available: \url{https://onnx.ai/}
\BIBentrySTDinterwordspacing

\bibitem{wenger2021backdoor}
E.~Wenger, J.~Passananti, A.~N. Bhagoji, Y.~Yao, H.~Zheng, and B.~Y. Zhao, ``Backdoor attacks against deep learning systems in the physical world,'' in \emph{Proceedings of the IEEE/CVF conference on computer vision and pattern recognition}, 2021, pp. 6206--6215.

\bibitem{zheng2022concealed}
J.~Zheng, P.~P. Chan, H.~Chi, and Z.~He, ``A concealed poisoning attack to reduce deep neural networks’ robustness against adversarial samples,'' \emph{Information Sciences}, vol. 615, pp. 758--773, 2022.

\end{thebibliography}

\end{document}